# Vibrational Frequencies of Cerium Oxide-Bound CO: A Challenge for Conventional DFT Methods

Pablo G. Lustemberg,[1,2] Philipp Pleßow,[3] Yuemin Wang,[4] Chengwu Yang,[4] Alexei Nefedov,[4] Felix Studt,[3] Christof Wöll,[4,*] M. Verónica Ganduglia-Pirovano[2,*]

[1]*Institute of Physics Rosario, IFIR, National Scientific and Technical Research Council, CONICET, and National University of Rosario, UNR, S2000EKF Rosario, Santa Fe, Argentina*

[2]*Institute of Catalysis and Petrochemistry, ICP, Spanish National Research Council, CSIC, 28049 Madrid, Spain*

[3]*Institute of Catalysis Research and Technology, IKFT, Karlsruhe Institute of Technology, KIT, 76344 Eggenstein-Leopoldshafen, Germany*

[4]*Institute of Functional Interfaces, IFG, Karlsruhe Institute of Technology, KIT, 76344 Eggenstein-Leopoldshafen, Germany*

In ceria-based catalysis, the shape of the catalyst particle, which determines the exposed crystal facets, profoundly affects its reactivity. The vibrational frequency of adsorbed carbon monoxide (CO) can be used as a sensitive probe to identify the exposed surface facets, provided reference data on well-defined single crystal surfaces together with a definitive theoretical assignment exist. We investigate the adsorption of CO on the $CeO_2$(110 and (111) surfaces and show that the commonly applied DFT(PBE)+$U$ method does not provide reliable CO vibrational frequencies by comparing with state-of-the-art infrared spectroscopy experiments for monocrystalline $CeO_2$ surfaces. Good agreement requires the hybrid DFT approach with the HSE06 functional. The failure of conventional DFT is explained in terms of its inability to accurately describe the facet- and configuration-specific donation and backdonation effects that control the changes in the C–O bond length upon CO adsorption and the CO force constant. Our findings thus provide a theoretical basis for the detailed interpretation of experiments and open up the path to characterize more complex scenarios, including oxygen vacancies and metal adatoms.

[*]Corresponding author
christof.woell@kit.edu, vgp@icp.csic.es

Cerium dioxide ($CeO_2$ or ceria) is extremely important in catalysis, either as a catalyst or as a support material [1], as well as in many other applications such as oxide-ion conductors in solid-oxide fuel cells [2] and lately in biology [3,4]. It is known that the ceria surface structure may alter the catalytic activity of ceria-based catalysts [5], and thus, ceria nanocrystals with controlled morphologies, such as nanooctahedra ({111} facet), nanocubes ({100} facet), and nanorods ({100} and {110} facets) are fabricated [6], and their chemisorption and reactivity properties investigated [7-12]. However, under reaction conditions ceria facets can undergo restructuring [13-17]. These findings make it crucial to develop characterization methods that are able to probe the surface of ceria-based catalysts with high specificity under operando conditions [18]. The characterization of oxide surfaces via a characteristic shift of the vibrational band of small probe molecules has high potential in this regard but crucially depends on highly accurate reference data.

CO is –in principle– a useful probe molecule and thus a fairly large number of infrared spectroscopy (IR) studies for ceria powder samples exposed to CO have been reported [19-23]. However, for a reliable assignment of CO vibrational frequencies, as required for example in the surface-ligand IR (SLIR) approach to characterize oxide particles [24], reference data for CO adsorbed on well-defined single crystal surfaces together with a definitive theoretical assignment are required. Unfortunately, theoretical methods, largely depending on density functional theory (DFT), have been lacking behind in the accurate description of the small shift of the C–O stretching frequency induced by its weak interaction with the oxide. This has been the motivation for pursuing experimental and theoretical research on the adsorption of CO onto extended monocrystalline ceria surfaces [25-31]. As of yet, studies that combine computational modelling and experiment are limited to CO adsorption on the (111) surface [25, 26]. For one monolayer (ML) CO adsorption on the oxidized $CeO_2$(111) surface, there is a blue shift ($\Delta\upsilon$) of +11 cm$^{-1}$ with respect to the CO vibrational stretching frequency in the gas phase [25], a common scenario for CO adsorbed on metal oxides [32]. The latest reported theoretical $\Delta\upsilon$ value of +9 cm$^{-1}$ [26] is in good agreement with experiment; it has been obtained by employing dispersion-corrected density-functional theory (DFT) in the DFT(PBE)+$U$(4.5) approach [33], with the PBE (Perdew-Burke-Ernzerhof) functional [34] and an effective Hubbard $U$-like term of 4.5 eV added for the Ce 4$f$ states [35]. However, this apparently good agreement is limited to a very diluted CO adlayer (1/16 ML), since for full coverage, the agreement is less satisfying, as we will show below.

The DFT(PBE) approach, which is based on the generalized gradient approximation (GGA) for the exchange-correlation energy, is known to badly fail in accurately describing reduced ceria [36-38], whereas DFT(PBE)+$U$, is commonly applied in the study of ceria-based systems, but questions regarding the best value for the $U$ parameter are still under debate [35, 38-42]. However, there is consensus that hybrid functionals using mixtures of DFT(GGA) and Hartree-Fock exchange energies (e.g., the HSE06 functional [43]) yield higher computational accuracy for ceria-based systems [38, 42, 44, 45, 46, 47]. Whether hybrid DFT is able to resolve the problems in describing CO adsorption on differently oriented ceria surfaces described above, that is the question. The use of hybrid functionals has rarely been considered crucial for the description of oxide-bound CO, with the exception of CO on MgO(001) [48,49], and we recall that hybrid DFT has helped resolve the CO/Pt(111) puzzle associated with the failure of DFT(GGA) approaches to properly describe CO-metal interactions [50].

In this Letter, we present a detailed analysis of CO adsorbed on the $CeO_2$(111) and (110) surfaces and demonstrate that the widely used DFT(PBE)+$U$ approach is not well suited for the prediction of CO vibrational frequencies. While being computationally fairly inexpensive, in some cases this method even is in qualitative disagreement with experiment, *i.e.*, instead of the typical blue shift of CO on oxides, a red shift is predicted. We provide firm computational evidence that for CO adsorbed on ceria surfaces the use of hybrid functionals is mandatory. It is only at this level of theory that the experimentally observed larger blue shift for $CeO_2$(110) by 17 cm$^{-1}$ [15] compared to $CeO_2$(111), is recovered. A careful analysis allows to relate the failure of the simpler DFT(PBE)+$U$ approach to the inability of the PBE functional to provide an accurate description of details of the electronic structure, in particular the facet- and configuration-specific donation and backdonation effects that control the changes in the C–O bond length upon CO adsorption and the CO force constant.

Moreover, the combination of high-resolution *p*-polarized IRRAS data with the results of DFT(HSE06) calculations allows us to gain unique insight into the CO adsorption modes on ceria surfaces. In Fig. 1 we show the results of IRRAS experiments of 1 ML adsorbed CO on single crystalline $CeO_2$(111) and $CeO_2$(110) surfaces. For $CeO_2$(111) a single intense band is visible for *p*-polarization at 2154 cm$^{-1}$ (Fig. 1, red circles), blue shifted $\Delta\upsilon$= +11 cm$^{-1}$ relative to the gas phase (2143 cm$^{-1}$ [51]). In previous work, this band has been assigned to CO bound to $Ce^{4+}$ cations embedded in a perfect single-crystalline $CeO_2$(111) surface environment [25]. For the $CeO_2$(110) surface, the CO SLIR data (Fig. 1, blue circles), reveal a more complex situation. For the saturated surface, a strong negative band is seen for *p*-polarization at 2071 cm$^{-1}$, accompanied by a smaller positive feature at 2160 cm$^{-1}$. Clearly, the data for the (110) surface are not consistent with just one peak. In fact, a fit using a negative band at 2171 cm$^{-1}$ and a positive band at 2160 cm$^{-1}$ (Fig. 1, upper panel, black line) yields good agreement with the data. In earlier work [15], the higher frequency peak (2171 cm$^{-1}$) has been tentatively assigned to CO bound to $Ce^{4+}$ cations, while the weaker peak at 2160 cm$^{-1}$, indicating the presence of a second species, has not been discussed. The sign of the feature (positive rather than negative) indicates the presence of a strongly tilted species with a substantial component of the transition dipole moment oriented parallel to the substrate.

Apart from CO on MgO(001) [49]], the presence of a strongly tilted CO species at a cationic site on a low indexed single crystal oxide surface is uncommon [52]. In addition, it is surprising that the vibrational frequency of CO bound to $Ce^{4+}$ sites differs by almost 20 cm$^{-1}$ between (111) and (110) surfaces – in many previous works it has been assumed that these frequencies are mainly sensitive to the charge state of the metal ion; see [24] and references therein. A thorough understanding of the binding of small reactants such as CO on oxide surfaces is of fundamental interest and essential if CO is used as an IR probe molecule for determination of exposed surface facets.

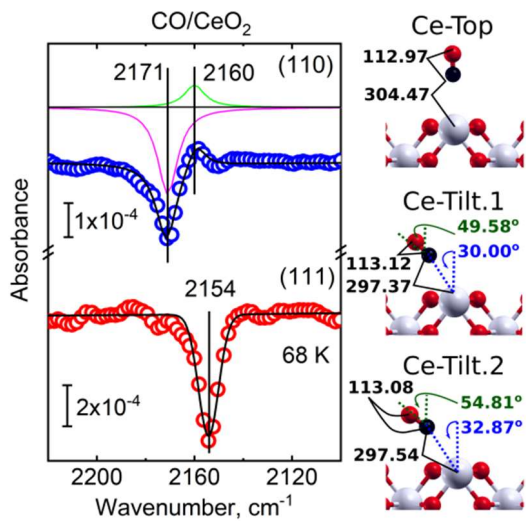

**Figure 1.** IRRAS spectra of 1ML CO adsorption on oxidize $CeO_2(111)$ (red circles) and $CeO_2(110)$ (blue circles) surfaces with *p*-polarized light as well as HSE06 optimized CO-$CeO_2(110)$ structures. Selected interatomic distances (in pm) are indicated. Black lines show the corresponding fits. The upper panel presents two-peaks deconvolution, corresponding to bands at 2171 cm$^{-1}$ (magenta) and 2160 cm$^{-1}$ (green). Adapted with permission from [25], Copyright 2014 Royal Society of Chemistry and with permission from [15], Copyright 2017 John Wiley & Sons, Inc.

We have first analyzed the binding of 1 ML CO on the $CeO_2(110)$ and (111) surfaces by performing high-precision DFT(PBE)+$U$(4.5) calculations (see Supporting Information). To our surprise, for the most strongly bound CO species on the $CeO_2(110)$ surface, instead of the typical blue shift of CO adsorbed on metal oxides, the results revealed a distinct (−23 cm$^{-1}$) red shift (Table 1), and for CO adsorbed on $CeO_2(111)$, as already mentioned, the shift is about zero (+1 cm$^{-1}$), in pronounced contrast to experimental observations (Fig. 1). At this point, we must come to the conclusion that theory at the DFT(PBE)+$U$ level is not adequate to describe the bonding situation of CO on ceria surfaces.

**Table 1.** Calculated DFT energies for the binding of 1ML CO to the CeO$_2$(111) and (110) surfaces with (1×1) periodicity, scaled CO stretching vibrational frequency (by the method-dependent factor $\lambda = \upsilon_{CO_{gas}}^{exp}/\upsilon_{CO_{gas}}^{calc}$ with $\upsilon_{CO_{gas}}^{exp}$ = 2143 cm$^{-1}$ [51] and $\upsilon_{CO_{gas}}^{calc}$ as given in Table S2), frequency shift, Δυ, and change in the C–O bond length, Δ$d_{C-O}$, upon adsorption with respect to the calculated values for the gas phase molecule, respectively. $d_{Ce-C}$ is the distance between the C and the Ce$^{4+}$ site to which the CO is attached.

| Method | Site | E$_{ads}$ (eV) | $d_{Ce-C}$ (pm) | Δ$d_{C-O}$ (pm) | υ (cm$^{-1}$) | Δυ (cm$^{-1}$) |
|---|---|---|---|---|---|---|
| (111) | | | | | | |
| IRRAS | | | | | 2154 | +11 |
| PBE+$U$(4.5) | atop | −0.03 | 288 | +0.05 | 2144 | +1 |
| HSE06 | atop | −0.09 | 295 | −0.06 | 2157 | +14 |
| (110) | | | | | | |
| IRRAS | | | | | 2171/2160 | +28/+17 |
| PBE+$U$(4.5) | atop | −0.16 | 299 | −0.12 | 2151 | +8 |
| | tilt.1 | −0.22 | 290 | +0.09 | 2120 | −23 |
| | tilt.2 | −0.21 | 291 | +0.06 | 2130 | −13 |
| HSE06 | atop | −0.15 | 304 | −0.18 | 2164 | +21 |
| | tilt.1 | −0.20 | 297 | −0.03 | 2145 | +2 |
| | tilt.2 | −0.19 | 298 | −0.07 | 2150 | +7 |

In order to resolve this issue, we have conducted a systematic study employing different levels of theory and in particular using hybrid functionals. We have first applied DFT with a number of semi-local GGA functionals, namely, BEEF-vdW [53], PBE [34], and PBEsol [54], as well as the DFT(PBE)+$U$(4.5) method to 1 monolayer (ML) CO adsorbed on the fully-oxidized CeO$_2$(111) surface modeled by (1 × 1) surface unit cells (see the Supporting Information for further details). Moreover, we have also employed the nonlocal HSE06 hybrid functional [43]. Additionally, the D3 dispersion correction (DFT+D3) [55,56] has been used in conjunction with the PBE, PBEsol, and HSE06 functionals, as well as with the PBE+U approach. The calculated CO adsorption energies ≲ 0.3 eV (in absolute value) with the GGA and GGA+$U$ methods (Tables 1 and S3) are consistent with those of previous PBE/PW91 [28, 29], PBE/PW91+U(4.5/5) [27, 28], and PBE+$U$(4.5)+D3 analyses. The HSE06 adsorption energy is similar to that calculated with PBE+$U$(4.5) within less than 0.1 eV (Table 1).

Independent of the employed DFT method, the most energetically preferable adsorption site on the $CeO_2$(111) surface is atop $Ce^{4+}$ with the C–O bond (in most cases) perpendicular to the surface. Interestingly, the distance of the C atom to the $Ce^{4+}$, $d_{Ce-C}$, and the change of the C–O bond length upon adsorption with respect to the calculated value for the gas phase molecule, $\Delta d_{C-O}$, depend on the approximation to the exchange-correlation functional in DFT (Tables 1 and S3). The $d_{Ce-C}$ distance is predicted to be shorter with most GGA-type functionals compared to that obtained with the hybrid HSE06 functional. Moreover, we found the C–O bond lengthened by about 0.1 pm for all GGA-type functionals, but shortened for the hybrid HSE06 functional by about the same amount (Tables 1 and S3).

The CO stretching vibrational frequency computed with the GGA/GGA+$U$ approximations (Tables 1 and S3) are by 6–10 cm$^{-1}$ smaller than the experimental value of 2154 cm$^{-1}$ (Fig. 1) and the calculated blueshifts of +1 to +5 cm$^{-1}$ are systematically smaller than the experimental value of $\Delta\upsilon$= +11 cm$^{-1}$. However, the hybrid HSE06 functional predicts values of $\upsilon$= 2157 cm$^{-1}$ and $\Delta\upsilon$= +14 cm$^{-1}$ that are in good agreement with the experiment (Table 1).

On the (110) surface, a similar adsorption site for CO atop of a $Ce^{4+}$ ion is found with the C–O bond nearly perpendicular to the surface, as well as two strongly tilted configurations (Fig. 1). The difference between the tilted configurations is that in one case the CO molecule points to the $Ce^{4+}$ ion in the second oxide plane (tilt.1, Fig. S1c, g) and in the other one to the $Ce^{4+}$ ion in the surface along the [1-10] direction (tilt.2, Fig. S1d, h). In the literature [27-29], in addition to the atop $Ce^{4+}$ site, the formation of strongly bound carbonate $CO_3^{2-}$–like species and the reduction of the ceria surface have been reported, but such species have not been observed in the IRRAS experiment [15]. The PBE+$U$(4.5) calculated binding energy for the atop $Ce^{4+}$ site $\lessapprox$ 0.2 eV (in absolute value, Table 1) is consistent with that of previous PBE/PW91+$U$(4.5/5) [27, 28] and PBE/PW91 [28, 29] studies. In the two tilted configurations, the adsorption energy is practically the same and is by less than 0.1 eV larger than in the atop one (Table 1). Once again, the HSE06 calculated binding energies are comparable to those obtained with PBE+$U$(4.5).

As mentioned above, for CO adsorption on the $CeO_2$(110) surface two blue shifted peaks by $\Delta\upsilon$= +28 and +17 cm$^{-1}$ are observed (Fig. 1). Inspection of Table 1 reveals that similar to the case of the (111) surface, for the (110), the deviations of the computed $\upsilon$ and $\Delta\upsilon$ values with PBE+$U$(4.5) from the experimental ones are substantial and that only the HSE06 values are in good agreement with the results of the IRRAS experiment with deviations $\lessapprox$ 9 cm$^{-1}$. The more intense band at 2171 cm$^{-1}$ and the less intense one at 2160 cm$^{-1}$ are here assigned to CO bound to $Ce^{4+}$ in atop and tilted configurations, respectively.

Inspection of the structures for the adsorption of CO on the CeO$_2$(110) surfaces reveals that the Ce–C distances are shorter with PBE+$U$ than with HSE06 (Table 1, Fig. S1), which is similar to the case of the CeO$_2$(111) surface. Also, the C–O bond length is longer with PBE+$U$ than with HSE06, and the corresponding changes in the C–O bond length, $\Delta d_{C-O}$, upon adsorption show that (in most cases) $\Delta d_{C-O} > 0$ with PBE+$U$ and $\Delta d_{C-O} < 0$ with HSE06 (Table 1). Specifically, the C–O bond is calculated to be compressed by 0.18 pm and up to about 0.1 pm in the atop and tilted configurations, respectively, by the HSE06 functional. However, for the tilted configurations, the PBE+$U$(4.5) calculated C–O bonds are stretched by about 0.1 pm. A compression of about 0.1 pm has been previously calculated for the atop Ce$^{4+}$ configuration employing the hybrid B3LYP functional [30,31].

In summary, with the PBE+$U$ functional the agreement of theoretical vibrational frequencies with the IRRAS experiment is rather poor, while with the HSE06 approach almost quantitative agreement can be achieved. To shed more light on this problem, we further inspected the density of states (DOS) for the CO adsorbed systems compared to that for the CO molecule in the gas phase. As an example, we focus here on the atop Ce configuration on the (111) surface (Fig. 3 and Fig. S4). The electronic state reminiscent of the CO 5σ orbital is clearly perturbed upon adsorption (Fig. 2a, b). The shorter Ce–C distance obtained with PBE+$U$ compared to HSE06 relates to a stronger σ-donation type of interaction by which charge is donated to the empty Ce $d$-states, as evidenced by the more pronounced electron density accumulation at the Ce–C bond (cf. Figs. 2c, d). As a consequence of the σ-donation, a *shortening* of the C–O bond is expected due to bond depopulation which implies $\Delta d_{C-O} < 0$. However, as stated above, it is only with the HSE06 functional that the computed $\Delta d_{C-O}$ values are negative, and those obtained with PBE+$U$ are positive.

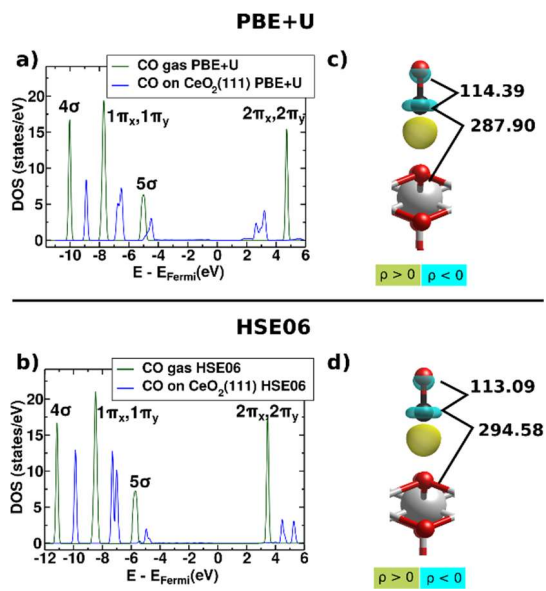

**Figure 2.** Total density of states (DOS) projected onto the CO (blue) for 1ML CO adsorbed on the $CeO_2(111)$ surface in the atop $Ce^{4+}$ configuration with the (a) PBE+$U$(4.5) and (c) HSE06 approaches compared to that of CO in the gas phase. (b) and (d) Isosurfaces of the charge density difference, namely, that of the CO adsorbed system from which both the charge density of the clean $CeO_2(111)$ surface (with a structure corresponding to that of the adsorbed system) and that of the layer of adsorbed CO molecules have been subtracted. Selected interatomic distances (in pm) are indicated.

In addition, the PBE+$U$ derived charge density difference upon CO adsorption on the $CeO_2(111)$ surface (Figs. 2c) reveals that electron density on the O atom of the molecule is enhanced, indicating that antibonding orbitals of CO have been populated. In the HSE06 calculations this electron density enhancement on the O end is less noticeable (Fig. 2d). As a result of the surface → CO backdonation type of interaction, a *lengthening* of the C–O bond is expected due to bond population. The overall change in the C–O bond upon adsorption is the result of synergistic charge transfer effects, namely, CO → surface donation and surface → CO backdonation, which cause the shortening and lengthening of the C–O bond length respectively. The PBE+$U$ approach predicts larger CO → surface and surface → CO charge transfers compared to HSE06 and as a result of which $\Delta d_{C-O} > 0$. Similar findings were obtained for 1ML CO adsorption on the $CeO_2(110)$ surface (Figs. S5 and S6). The 0.25 ML CO adsorption on both surfaces has also been considered (see Figs. S2 and S3).

The calculated CO vibrational frequency of the gas phase molecule as a function of a varying C–O bond length shows a clear inverse correlation (Fig. S7). The shorter the C–O bond, the higher the vibrational CO stretching frequency leading to a blueshift, whereas the longer the C–O bond, the lower the frequency, resulting in a redshift. The HSE06 computed $\Delta \upsilon$ and $\Delta d_{C-O}$ changes upon CO adsorption on the $CeO_2(111)$ and (110) surfaces, qualitatively follow the inverse correlation. However, with PBE+$U$ deviations are significant.

Having established the ability of the HSE06 functional to provide reliable CO vibrational frequency shifts of CO adsorbed on ceria surfaces upon comparison with the experimental IRRAS data (selected calculations using the PBE0 [57] functional give similar results, see Supporting Information), we note the important result that the shifts are facet- and configuration-dependent, because the final changes in the C–O bond length upon adsorption depend on them too, and thus CO can be used as a sensitive chemical probe molecule in the characterization of the specific facets exposed by ceria catalysts. Specifically, for the atop configurations on the (110) and (111) surfaces, blueshifts of the CO stretching frequency of +21 and +14 cm$^{-1}$, respectively, have been computed and related to a shortening of the C–O bond of 0.18 and 0.06 pm, respectively. Moreover, an additional blueshifted frequency of +7 cm$^{-1}$ is calculated for the (110) surface and related to the shortening of 0.07 pm of the C–O bond of a tilted adsorbed CO. Facet-specific donation and backdonation effects were revealed no only on $CeO_2$, as demonstrated in this work, but also on $TiO_2$ anatase nanoparticles exposing the {101} and {110} facets [58].

Generally speaking, the characterization of metal oxide nanoparticles under realistic conditions is of great importance for catalytic applications and is paramount for exploiting the interplay between structure and reactivity. A rather large variety of different vibrational frequencies can be observed for powder catalysts when using IR probe molecules, and a solid basis for a thorough assignment is needed. DFT calculations validated by IRRAS data for well-defined single crystals can provide such a reliable reference. The validation process here described, demonstrated a failure of the frequently used GGA functionals in DFT, for the example of $CeO_2$-bound CO. As a consequence, earlier interpretations of measured spectra based on GGA calculations may need to be revised. Importantly, the HSE06 approach provides the required accuracy and allows reproducing the facet- and configuration-dependent vibrational spectra for CO adsorbed on the $CeO_2$(111) and (110) surfaces. This high accuracy achieved for one of the most challenging oxide surface manifests a striking advancement in the theoretical description of metal-oxides and opens up the path for analyzing more complex scenarios in future work, including oxygen vacancies and metal adatoms.

**Acknowledgements**

This project received funding from the European Union's Horizon 2020 research and innovation programme under the Marie Skłodowska-Curie grant agreement No 832121. Computer time provided by the BIFI-ZCAM and the RES (Red Española de Supercomputación) resources at MareNostrum 4 (BSC, Barcelona) and Altamira (IFCA, Cantabria) nodes, as well as by the DECI resources at Finis Terrae II based in Spain at CESGA, with the support from PRACE aislb, is acknowledge. M.V.G.P. thanks the support by the MICINN-Spain (RTI2018-101604-B-I00). The DFT data that support the findings of this study are available in Materials Cloud {https://www.materialscloud.org/home} with the identifier doi: 10.24435/materialscloud:z7-z0.